# Single-core Or Multi-core? A Mini Review on Magnetic Nanoparticles for Magnetic Particle Spectroscopy-based Bioassays

Kai Wu[†,*], Jinming Liu[†], Diqing Su[‡], Renata Saha[†], Vinit Kumar Chugh[†], and Jian-Ping Wang[†,*]
[†]Department of Electrical and Computer Engineering, University of Minnesota, Minneapolis, Minnesota 55455, USA
[‡]Department of Chemical Engineering and Material Science, University of Minnesota, Minneapolis, Minnesota 55455, USA
*E-Mails: wuxx0803@umn.edu (K.W.), jpwang@umn.edu (J.-P.W.).

**Abstract**
Magnetic particle spectroscopy (MPS) is a technology that derives from magnetic particle imaging (MPI) and thrives as a standalone platform for many biological and biomedical applications, benefiting from the facile preparation and chemical modification of magnetic nanoparticles (MNPs). In recent years, MPS has been reported in extensive literatures as a versatile platform for different bioassay purposes using artificially designed MNPs, where the MNPs serve as magnetic tracers, the surface functionalized reagents (e.g., antibodies, aptamers, peptides, etc.) and tiny probes capturing target analytes from biofluid samples. The biochemical complexes on MNP surfaces can be tailored for different bioassay requirements, while the design of MNPs are of less attention for MPS-based bioassays. For MNPs in most bioassay applications, superparamagnetism is prerequisite to avoid agglomerates and false magnetic signals. Single- and multi-core superparamagnetic nanoparticles (SPMNPs) are prevalently used in MPS-based bioassays. In this mini review, we compared the pros & cons of different MPS platforms realizing volumetric- and surface-based bioassays with single- and multi-core nanoparticles, respectively.

## 1. Introduction

The application of magnetic nanoparticles (MNPs) in biological and biomedical context is a fast-growing area. Where MNPs, with proper surface functionalization, are used as contrast agents in magnetic resonance imaging (MRI) and nuclear magnetic resonance (NMR)-based biosensors[1–6], tracers for magnetic particle imaging (MPI) and magnetic particle spectroscopy (MPS)-based biosensors[7–13], tags for magnetoresistive (MR)-based biosensors[14–19], labels for cell sorting and separation[20,21], heating sources for hyperthermia[22–24], carriers for drug/gene delivery[22,23], and surface-enhanced Raman spectroscopy (SERS)-active substrate for different biochemical assays[25–27]. Iron oxide nanoparticles (IONs, e.g., $Fe_3O_4$ and $\gamma$-$Fe_2O_3$) are frequently used for these applications due to high biocompatibility, stability, and biodegradability. Furthermore, the quest for high moment MNPs using other magnetic materials such as pure metals (e.g., Fe, Co, Ni), alloys (e.g., FeCo, alnico, permalloy), and oxides (e.g., $MFe_2O_4$ where M = Fe, Co, Mn, Ni, Zn)[28–30] is growing in view of higher magnetic signal and magnetic force for those aforementioned applications.

Besides choosing different MNP materials with high saturation magnetizations, an alternative to achieve high moment MNPs is to use larger magnetic cores. With increasing magnetic core size, these MNPs show ferrimagnetic behavior and hysteresis loops (non-zero remanent magnetization), which, is preferred for some applications such as magnetic hyperthermia. However, this remanent magnetization leads to agglomeration of MNPs, blocking blood vessels and causing false magnetic signals for imaging and biosensing[31–33]. Thus, MNPs with superparamagnetic behaviors (zero remanent magnetization) are chosen for those applications. Superparamagnetism appears in very small MNPs when the magnetizations randomly flip due to thermal fluctuations. The core sizes of superparamagnetic nanoparticles (SPMNPs) are usually below several nanometers to several tens of nanometers depending on the materials[34]. Consequently, the requirement on magnetic core size for superparamagnetism limits the achieving of high magnetic moment SPMNPs. Thus, multi-core SPMNPs are designed where a cluster of smaller SPMNPs are embedded in a polymer matrix[35–37]. These relatively large, multi-core SPMNPs show negligible remanent magnetization (compared to single-core MNPs with the same overall size), higher colloidal stability, and low tendency to form agglomerates, which is an excellent alternative to single-core SPMNPs. Multi-core SPMNPs have also been exploited for magnetic separation[38,39], MPI[36,40,41], MRI[42–44], and hyperthermia[45]. While for most biological and biomedical applications, single- and multi-core SPMNPs hold equally important roles[46–48]. However, in the area of MPS-based bioassays, single-core SPMNPs are prevalently used and reported in most of the research articles[10,12,49–54]. Herein, we comment on the mechanism of MPS-based bioassays, the motivation of choosing single-core SPMNPs as tracers for traditional MPS-based bioassays. In addition, we summarize the specific scenarios where single- and/or multi-core SPMNPs are used.

## 2. Mechanism of MPS-based Bioassays

Upon the application of sinusoidal magnetic fields, the magnetic moments of SPMNPs in aqueous medium try to align with the fields through the joint effects of Brownian and Néel relaxation processes, to minimize magnetostatics energy, which are countered by the thermal fluctuations ($k_BT$, where $k_B$ is the Boltzmann constant and T is temperature). Néel process is the relaxation of magnetic moment inside a stational SPMNP to align with external magnetic field and there is no physical movement related to this process. While, Brownian process is the rotational movement of the whole SPMNP along with its magnetic moment in response to external magnetic field. In addition to thermal energy that randomizes magnetic moments, Brownian process is also affected by the hydrodynamic volume, $V_h$, of SPMNP as well as the viscosity, $\eta$, of the aqueous medium. On the other hand, Néel process is also affected by the effective anisotropy (including crystal and shape anisotropies), $K_{eff}$, and volume of magnetic core, $V_c$. These two processes jointly govern the ability of magnetic moments to follow the time-changing magnetic field, thus, tuning the dynamic magnetic responses to the field. The effective relaxation time is dominated by the faster relaxation process while for single-core SPMNP with magnetic core size above 25 nm (some papers and books reported this number to between 12 nm and 20 nm), the Brownian process will dominate dynamic magnetic responses of SPMNPs[55–59].

From a macro perspective, nonlinear dynamic magnetic responses of SPMNPs to external sinusoidal magnetic fields induce higher odd harmonics, which, are recorded by a pair of specially designed pick-up coils and converted into spectral components[10,53,54]. These higher harmonics are used as metrics for measuring the temperature (T) and viscosity ($\eta$) of SPMNP medium[55,60–64], as well as the saturation magnetization ($M_s$), magnetic core size ($V_c$), and hydrodynamic size ($V_h$) of SPMNP[49,65]. Since the Brownian process can be altered by the hydrodynamic size of SPMNPs, researchers have been exploiting the Brownian process-dominated SPMNPs for MPS-based bioassays. MPS-based bioassays detect target analytes (chemicals/biological compounds) from aqueous medium where the odd harmonics of SPMNPs are used as metrics for monitoring their hydrodynamic size ($V_h$), which, reflects the bound states of target analytes to SPMNPs[10]. Where the capture antibodies (or aptamers, peptides, etc.) are functionalized on SPMNP surface. Target analytes will bind to SPMNPs through specific antibody-antigen interactions (or DNA-DNA interaction, RNA-RNA interaction, etc.). This specific binding process allows us to quantify target analytes from the aqueous testing sample. The dynamic magnetic responses of SPMNPs will be impaired by the conjugation of target analytes and the Brownian process is countered. As a result, the observed harmonics become weaker and phase lag between magnetic moments to external field becomes larger[10,12,51,52]. The magnetic signals are recorded by the pick-up coils in the format of time-varying voltage signals, converting the quantity of target analytes to spectral components - harmonics. Nowadays, MPS has been actively explored as a portable, highly sensitive, low cost, and easy-to-use *in vitro* bioassay platform.

## 3. Single- and Multi-core SPMNPs as Tags for MPS-based Bioassays

In the foregoing section, we mentioned the two relaxation processes that govern the dynamic magnetic responses of SPMNPs and explained the reasons behind choosing Brownian process-dominated SPMNPs for MPS-based bioassays. For single-core SPMNPs suspended in aqueous medium, Brownian process-dominated SPMNPs could be artificially controlled and selected by tuning the size of particles. However, in multi-core SPMNPs, a cluster of smaller SPMNPs are embedded in the matrix where their Brownian process is blocked. The collective dynamic magnetic responses of multi-core SPMNPs and smaller SPMNPs embedded are governed by complicated Brownian and Néel processes. Currently, there is no theoretical models available for this kind of magnetic response yet. Although multi-core SPMNPs show many merits such as negligible remanent magnetizations and high magnetic moment per particle, the free rotational Brownian process of SPMNPs embedded in matrix are blocked. As a result, these multi-core SPMNPs are not applicable for MSP-based bioassays that utilizes the freedom of rotational motion (Brownian process) as metric of bound state.

Multi-core SPMNPs are limited by the fact that Brownian process is blocked, and this type of particles are rarely reported in MPS-based bioassays. However, the properties such as higher magnetic moment per particle and being superparamagnetic are so appealing that they are used in an altered form MPS-based bioassays[66–69]. For traditional MPS-based bioassays, single-core

SPMNPs are pre-functionalized with capture antibodies (or aptamers, peptides, etc.), then biofluid sample is added, the real-time specific binding is monitored by MPS platform from the whole SPMNP system, see **Figure 1(A)**. This homogenous, volumetric-based, one-step, wash-free sensing scheme allows for "press one button and get the result within a few minutes". The quantity/concentration of target analytes is evaluated by the change of magnetic responses (harmonics from MPS spectra) before and after the addition of biofluid sample.

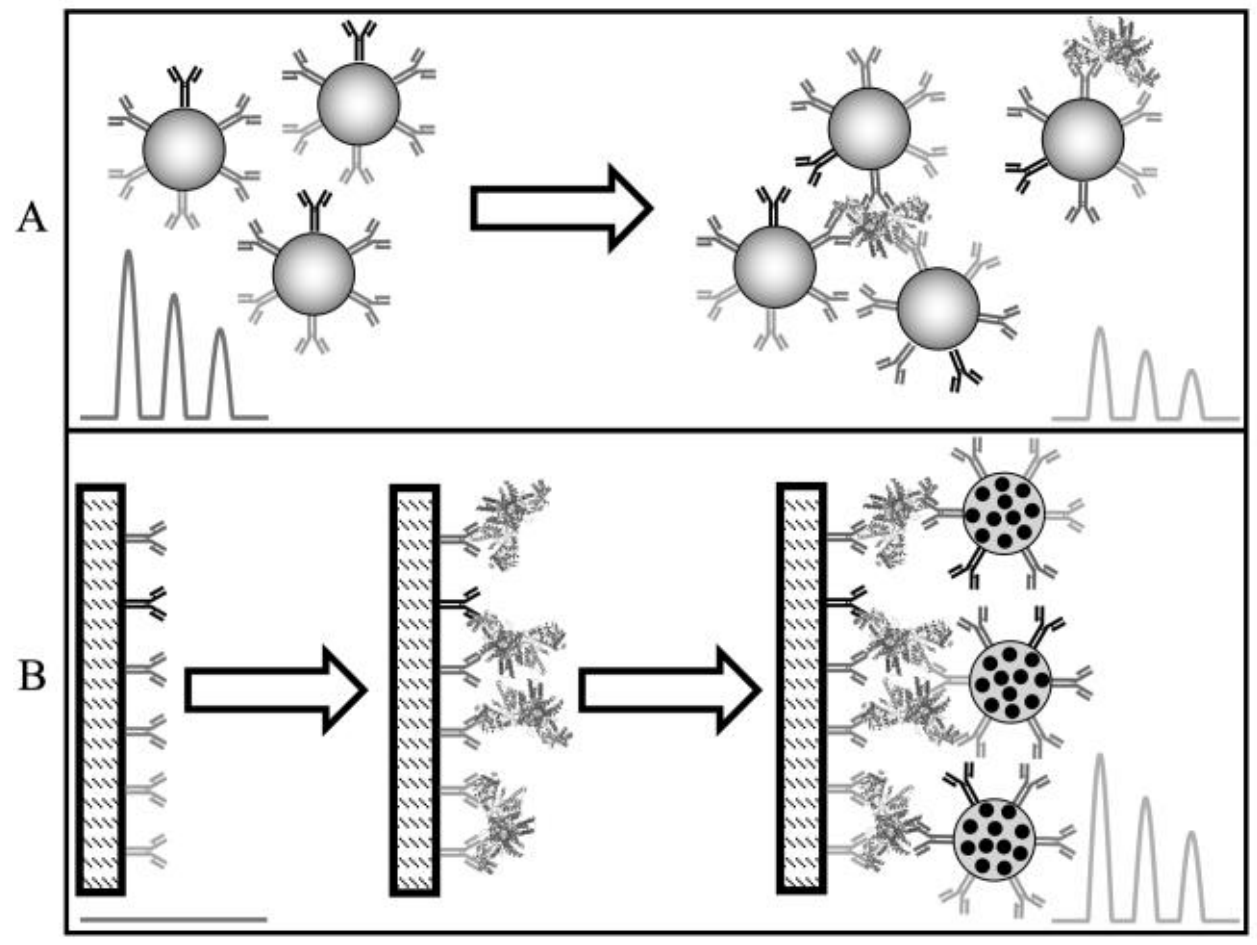

**Figure 1**. (A) Traditional MPS-based bioassay using single-core SPMNPs. Changes in odd harmonics from MPS spectra before and after adding samples as used as metrics. (B) An altered form MPS-based bioassay using multi-core SPMNPs. Harmonics from the captured of SPMNPs in the presence of target analytes are used as metrics.

On the other hand, multi-core SPMNPs have been reported as tracers in altered form MPS-based bioassays[69–72]. Although the complicated joint Brownian and Néel processes of multi-core SPMNPs are under investigation, the nonlinear dynamic magnetic responses can be used as metrics to indicate the capture of multi-core SPMNPs in the presence of target analytes. In an altered form MPS system, a nonmagnetic substrate functionalized with capture antibodies (or aptamers, peptides, etc.) provides a reaction surface and it allows the specific binding of target analytes from biofluid sample, see **Figure 1(B)**. Then the unbound biochemical compounds are removed by a flush wash step. Followed by passing detection antibody (or aptamer, etc.) functionalized multi-core SPMNPs though the surface. These multi-core SPMNPs bound to the immunocomplex on the reaction surface serve as magnetic labels to be recorded by the MPS system. An extra wash step to flush away unbound multi-core SPMNPs prior to MPS measurement is highly suggested to minimize detection error. Herein, the rotational freedom of SPMNPs is not used as metrics to reflect the bound states of target analytes. Instead, the nonlinear magnetic responses (due to superparamagnetism) of multi-core SPMNPs are recorded by MPS as metrics to quantify target analytes.

## 4. Conclusive Remarks

In summary, two types of MPS platforms, volumetric and surface-based, are reviewed. The volumetric-based MPS bioassay uses single-core SPMNPs and the detection of target analytes relies on the Brownian process of SPMNPs. While for multi-core SPMNPs where the Brownian process is blocked, they are used in surface-based MPS bioassay where the detection of target analytes relies on the dynamic magnetic responses of SPMNPs. Volumetric-based MPS bioassay utilizes the rotational freedom of single-core SPMNPs as metrics and the magnetic responses directly reflect the binding of target analytes from aqueous medium. Realizing "press one button and get the result within a few minutes". This detection mechanism allows for non-technicians with minimum training requirements to carry out self-testing in rural areas or at home.

On the other hand, multi-core SPMNPs can provide higher magnetic signal per particle over the single-core counterparts, thus, allows for better sensitivity bioassays. Although mutli-core sacrificed the rotational freedom of SPMNPs, surface-based MPS bioassay scheme utilizes the dynamic magnetic responses of multi-core SPMNPs. By adding a reaction surface and capturing these multi-core SPMNPs through specific antibody-antigen interactions (or DNA-DNA interaction, RNA-RNA interaction, etc.), the quantity of remaining multi-core SPMNPs after multiple wash steps are proportional to the number of target analytes from biofluid sample. Surface-based MPS with multi-core SPMNPs as tracers allow for highly sensitive detection while requires multiple wash steps. This limits its applications in lab contexts. The pros & cons of each platform are summarized and compared in **Table 1**.

**Table 1**. Comparison between single- and multi-core SPMNPs for MPS-based bioassays.

| | | | | | |
|---|---|---|---|---|---|
| **Single-core**[1] | Pros | · Allows one-step wash-free detection. | · Requires multiple wash steps to remove unbound biochemical compounds and SPMNPs. | Cons | **Multi-core**[2] |

| | | | |
|---|---|---|---|
| | · Low cost per test: the cost mainly come from SPMNP and capture antibodies (or aptamers, peptides, etc.). | · High cost per test: extra cost comes from nonmagnetic substrate and detection antibody (or aptamer, etc.). | |
| Cons | · Low magnetic signal per particle.<br>· Low sensitivity: homogenous, volumetric-based bioassay, magnetic signal comes from whole SPMNP sample. | · High magnetic signal per particle.<br>· High sensitivity: surface-based bioassay, magnetic signal only comes from those bound SPMNPs. | Pros |

[1]Single-core SPMNPs for volumetric-based MPS bioassay.

[2]Multi-core SPMNPs for surface-based MPS bioassay.

## Associated Content

**ORCID**

Kai Wu: 0000-0002-9444-6112

Jinming Liu: 0000-0002-4313-5816

Diqing Su: 0000-0002-5790-8744

Renata Saha: 0000-0002-0389-0083

Jian-Ping Wang: 0000-0003-2815-6624

**Notes**

The authors declare no conflict of interest.

**Acknowledgments**

This study was financially supported by the Distinguished McKnight University Professorship, the Centennial Chair Professorship, and the Robert F Hartmann Endowed Chair from the University of Minnesota.